\pdfoutput=1
\documentclass[12pt,a4paper]{article}%
\usepackage{amsmath,amssymb,amsfonts}
\usepackage{graphicx,epsfig,url}
\usepackage{color}
\usepackage{amsmath}
\usepackage{amsfonts}
\usepackage{amssymb}
\usepackage{float}
\usepackage{caption2}
\usepackage{graphicx}
\usepackage{hyperref}%
\numberwithin{equation}{section} 
\setcaptionmargin{1cm}

\begin{document}
\begin{titlepage}
\vskip 1.0cm
\begin{center}
{\Large \bf S-particles at their naturalness limits} \vskip 1.0cm {\large Riccardo
Barbieri$^a$\ \   and Duccio Pappadopulo$^b$} \\[1cm]
{\it $^a$ Scuola Normale Superiore and INFN, Piazza dei Cavalieri 7, 56126 Pisa, Italy} \\[5mm]
{\it $^b$ Institut de Th\'eorie des Ph\'enom\`enes Physiques, EPFL,  CH--1015 Lausanne, Switzerland}\vskip 1.0cm
\end{center}
\begin{abstract}
We draw attention on a particular configuration of supersymmetric particle masses, motivated by naturalness and flavour considerations.  All its relevant phenomenological properties for the LHC are described in terms of a few physical parameters, irrespective of the underlying theoretical model. This allows a simple characterization of   its main features, useful to define a strategy for its discovery.
\end{abstract}
\end{titlepage}



\section{A motivated configuration of s-particle masses}

Supersymmetric particles have been so far elusive. Nor we have seen any indirect signal of them in flavour physics, giving rise to the so-called \textit{supersymmetric flavour problem}. 
Assuming that s-particles at the Fermi scale indeed exist, is there any message in this? On one side, the discovery potential of the LHC in its full strength appears large enough to render this question not so urgent: if s-particles are there, they will be seen. On the other side, any orientation towards a specifically motivated configuration of s-particle masses and, possibly, of corresponding signals would clearly be valuable. This would be especially true in the case of particular configurations that appear to have received relatively little attention in the otherwise vast literature on phenomenological supersymmetry. We argue in the following that there is  one such case, based on naturalness and on the supersymmetric flavour problem. As we shall see, it is also simple to characterize some of its features, useful to define a strategy for its discovery.

Since naturalness is the main motivation for supersymmetry at the Fermi scale, it makes sense to ask which are the parameters affecting the s-particle spectrum that are mostly constrained by naturalness itself. We have in mind a generic Minimal Supersymmetric Standard Model, unconstrained by specific supersymmetry breaking mechanisms or specific unification conditions. The answer is well known: 
\begin{itemize}
\item The $\mu$-parameter, which affects the electroweak scale already at tree level and gives, in absence of mixing with the gauginos, the higgsino masses;
\item The masses $m_{\tilde{Q}}$ and $m_{\tilde{u}}$ of the third generation of squarks, which are coupled through the large top Yukawa coupling, $\lambda_t$, to the Higgs system and define, together with the stop-mixing term, the physical masses, $m_{\tilde t_1}, m_{\tilde t_2}$, of the two \textit{stops}, $\tilde{t}_1, \tilde{t}_2$, and of the left-handed \textit{sbottom}, $m_{\tilde b}$, in absence of mixing with the right-handed one.
\end{itemize}
These can be the lightest s-particle masses, whereas all the other s-partner masses, allowed to go to their naturalness limits, can be heavier or be phenomenologically (almost) irrelevant for LHC.  We find this possibility - to be made more precise in a while - motivated by two main considerations. If these are the s-particle masses, the 
supersymmetric flavour problem gets definitely alleviated, if not solved completely, depending on the assumptions one is willing to make on the (approximate) flavour symmetries. We have in mind, at least as an example, an approximate $U(2)$ symmetry which enforces CKM-like angles in the extra supersymmetric interactions and some amount of degeneracy between the s-fermions of the first two generations\cite{Pomarol:1995xc}\cite{Barbieri:1995uv}\cite{Barbieri:1997tu}. It is also true - without any bias from specific models - that both the $\mu$-parameter and the masses of the squarks coupled to the Higgs system via $\lambda_t$ are sufficiently singled out from the other mass parameters that one may conceive for them a somewhat special role. Here it suffices to mention the much discussed $\mu$-problem and, e.g., the effect of $\lambda_t$ on the running of $m_{\tilde{Q}}$ and $m_{\tilde{u}}$.

Let us be more specific on our choice of the mass parameters.
A relevant range for 
the $\mu$-parameter is between 100 and 200 GeV, to evade the LEP2 bound on charginos but otherwise not significantly above the Fermi scale itself, $v= 175$ GeV. The masses of $\tilde{t}_1, \tilde{t}_2$ and $\tilde{b}$ should  just be sufficiently heavy  to allow a Higgs boson mass again consistent with the LEP2 bounds: depending on the mixing term between the stops, their typical range is in the  hundreds of GeVs (see below).  The gluino mass, $m_{\tilde g}$, is also constrained by naturalness and flavour considerations. On the upper side, $m_{\tilde{g}}$ drives a one loop effect on $m_{\tilde{Q}}$ and $m_{\tilde{u}}$, that wants to make them too heavy, depending on the messenger scale. 
On the low side, gluino-sbottom exchanges in $\Delta F=2$ \cite{Barbieri:1997tu}\cite{Giudice:2008uk} and $\Delta B=1$ \cite{Barbieri:1993av}\cite{Haisch:2008ar} amplitudes with CKM-like angles point to a not too light gluino.
An interesting range for the gluino mass is between 500 GeV and 1 TeV. Naturalness allows  the remaining s-particles to be heavy enough 
to play little role in the LHC phenomenology, at least in a  first stage. This is the case for all the
s-fermions with small coupling to the Higgs system, that can go  in the $2\div 3$ TeV range\footnote{Unification considerations may suggest that not only $\tilde{t}_1, \tilde{t}_2$ and $\tilde{b}_L$ but all the third generation s-fermions be relatively lighter than those ones of the first two generations. This could still be consistent with current flavor constraints, although bringing the focus also on the $\mu\rightarrow e +\gamma$ transition, mediated by stau-gaugino exchanges, and on the ongoing experiment at PSI\cite{Bemporad:2008zz}. From the point of view of the LHC signals discussed below nothing would change as long as the right sbottom mass is comparable or heavier than $m_{\tilde{g}}$ and both $\tilde{\tau}$'s and the third generation s-neutrino(s) are heavier than $\mu$. In turn the heaviness of $\tilde{b}_R$ relative to $\tilde{b}_L$ could be due to the smallness of the bottom Yukawa coupling relative to the top one in their running effects on the squark masses.}\cite{Dimopoulos:1995mi},
and  for the electroweak gauginos, that can safely be taken at about 500 GeV or even higher. As we are going to see, an advantage of considering  this configuration of s-particle masses is that its main LHC phenomenology  is fully determined by a few physical parameters, irrespective of the underlying theoretical model.

\section{Relevant productions and decays at the LHC}

\begin{figure}[tb]
\centering
\includegraphics[width=9cm]{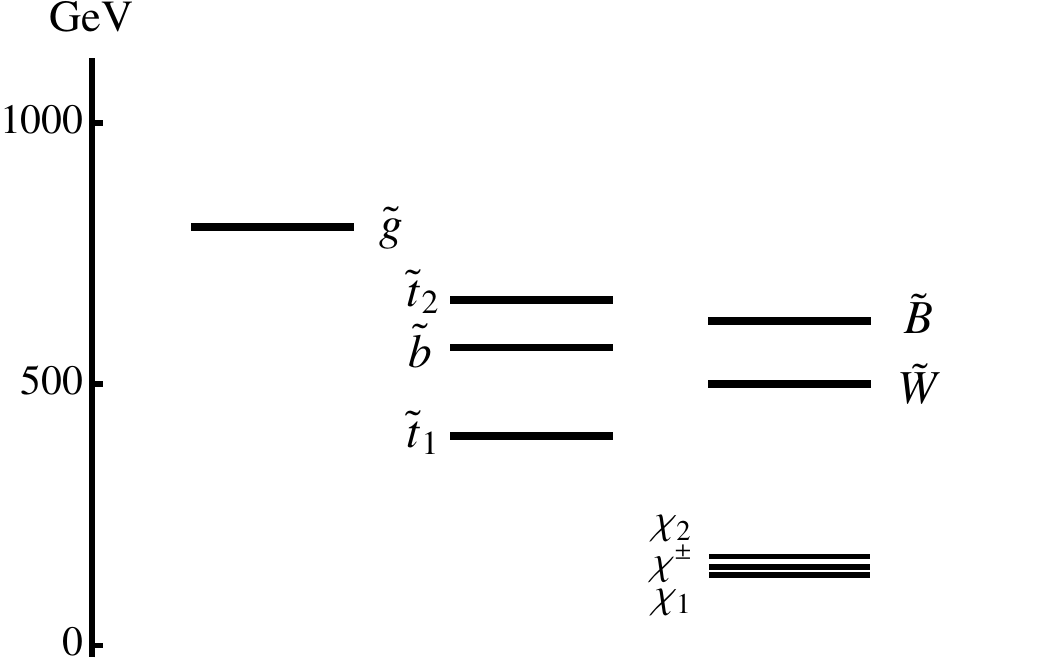} \caption{{\small A particular representation of the generic s-particle spectrum below 1 TeV considered in this paper}}
\label{spettro}
\end{figure}

Fig. \ref{spettro} shows a particular representation of the generic s-particle spectrum that we are led to consider below 1 TeV. The lightest s-fermions are mostly higgsino-like, have an average mass $\mu$, a small splitting among them, in the 10 GeV range, and a definite order, $m_{\chi_2} > m_{\chi^\pm} > m_{\chi_1} $. The decay among them of the heavier into the lighter ones via virtual $W$ and $Z$ produces also leptons, whose possible detection has been analyzed in Ref. \cite{Kitano:2006gv}. The softness of these leptons, however, make us consider other signals of this configuration of s-particles. As a consequence, the splitting among the light higgsinos can be safely neglected. Furthermore, as anticipated, the heavier (mostly) electroweak gauginos play no relevant role. Although they can occur in two  body decays of some squarks, the corresponding branching ratios are negligibly small,  at the relative {\it per cent} level, because of  the stronger coupling of  the squarks to the higgsinos  and because of phase space.

The most significant signals in the relatively earlier stages of  the LHC come from gluino pair production, followed, when allowed by phase space, by two-body gluino decays:
\begin{equation}
pp \rightarrow \tilde{g} \tilde{g};~~~ \tilde{g}\rightarrow \tilde{t}_1 \bar{t},~ \tilde{t}_2 \bar{t},~ \tilde{b} \bar{b}
\end{equation}
and the corresponding conjugate modes, each with equal branching ratios. All these branching ratios, given in Appendix A, are determined by $m_{\tilde g}, m_{\tilde t_1}, m_{\tilde t_2}$ and by the stop mixing angle $0<\theta_t<\frac{\pi}{2}$, defined in terms of the left and right stops by
\begin{equation}
\begin{pmatrix}
\tilde t_L\\
\tilde t_R
\end{pmatrix}=\begin{pmatrix}
\sin\theta_t && \cos\theta_t \\
-\cos\theta_t && \sin\theta_t 
\end{pmatrix} \begin{pmatrix}
\tilde t_1\\
\tilde t_2
\end{pmatrix}.
\end{equation}
The mass of the single relevant sbottom is\footnote{We are neglecting small terms in the squark mass-matrices squared proportional to $g^2 v^2$.}
\begin{equation}
m_{\tilde b}^2\approx \frac{m_{\tilde t_2}^2-m_{\tilde t_1}^2}{2} \cos 2 \theta_ t+\frac{m_{\tilde t_2}^2+m_{\tilde t_1}^2}{2}-m_t^2.
\end{equation}
By definition $m_{\tilde t_1} < m_{\tilde t_2}$.

\begin{figure}[tb]
\label{bsgamma}
\centering
\includegraphics[width=8cm]{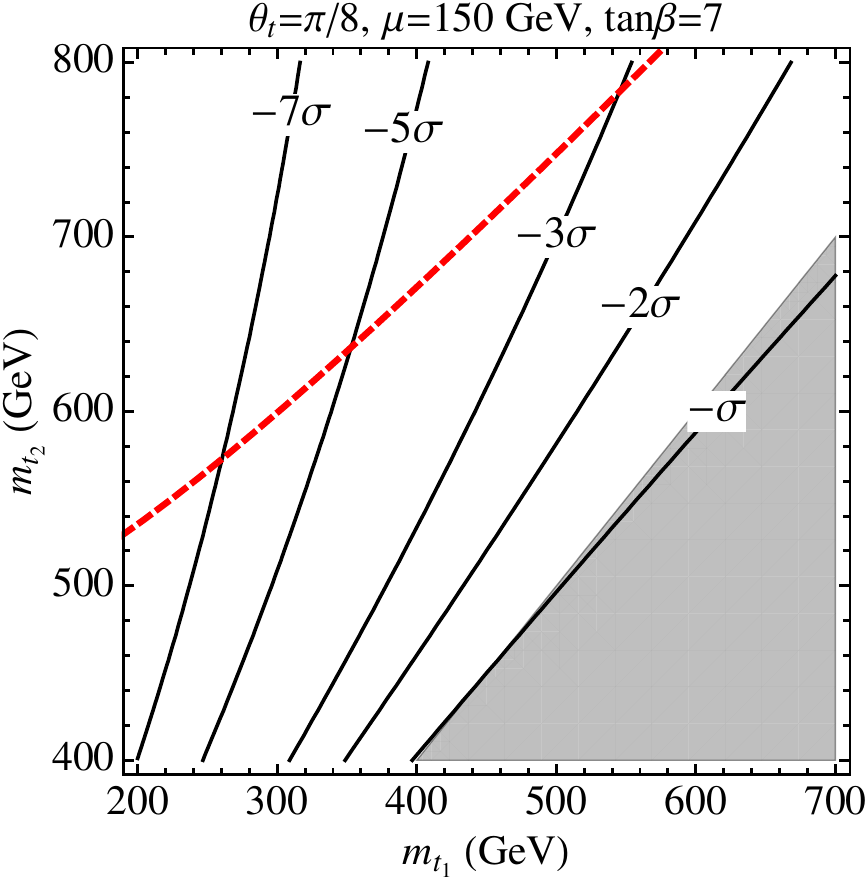} \includegraphics[width=8cm]{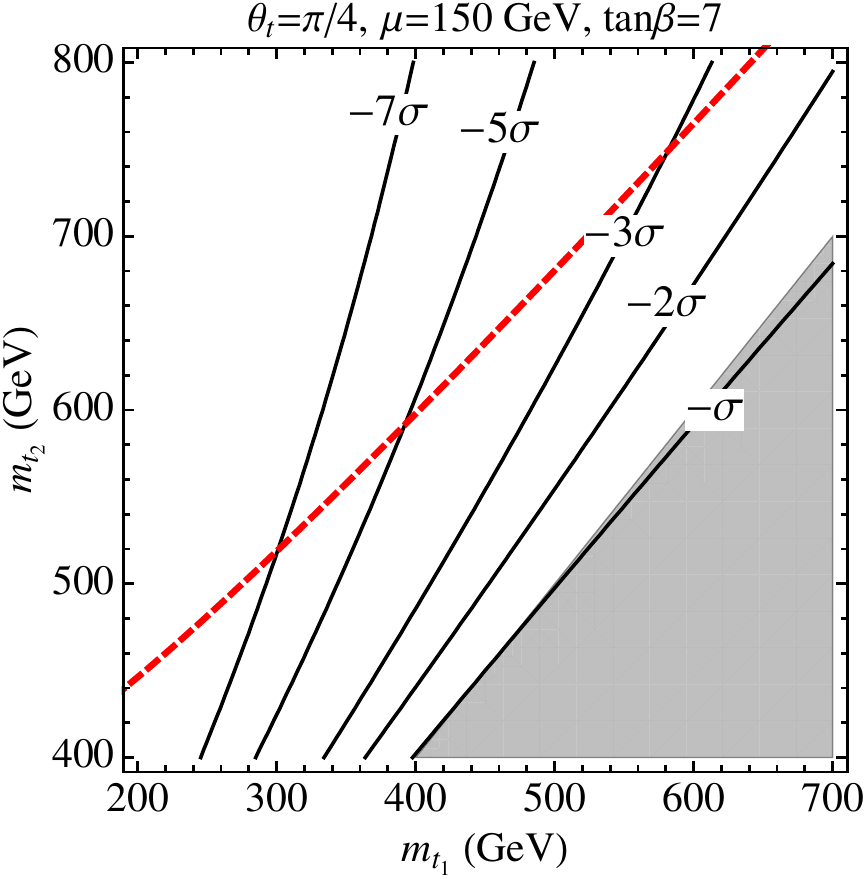} \caption{{\small Contours representing the deviation of the ratio $X=\frac{BR^{SM+SUSY}(b\rightarrow s\gamma)}{BR^{SM}(b\rightarrow s\gamma)}$ from its measured value $X^{EXP}=1.12\pm0.11$, in the illustrative case that the SUSY amplitude is {\it only} given by the higgsino-stop exchange with {\it exactly} CKM mixing angles. The LEP bound on the Higgs boson mass (see eq. (\ref{higgsmass}) in the Appendix) is satisfied above the red dashed line.
}}
\end{figure}

In a synthetic notation,  all the two-body decays of the three squarks, when allowed by phase space and charge conservation, are
\begin{equation}
\tilde{q} \rightarrow q \chi
\label{sqchi}
\end{equation}
\begin{equation}
\tilde{q} \rightarrow \tilde{q} V
\label{sqV}
\end{equation}
\begin{equation}
\tilde{q} \rightarrow  \tilde{q} S
\label{sqS}
\end{equation}
where
 \begin{equation}
\tilde{q} = \tilde{t}_1, \tilde{t}_2, \tilde{b};~~
q=t, b;~~
\chi = \chi_1, \chi_2, \chi^\pm;~~
V= W, Z;~~
S= h, A, H^\pm
\end{equation}
Other than $m_{\tilde t_1}, m_{\tilde t_2}$ and $\theta_t$, the branching ratios for these decays involve $\mu$ and $\tan{\beta}=v_2/v_1$, as is the case for the mass of the lightest scalar, $h$. All of these quantities are given in Appendix A. A relevant range for  $\tan{\beta}$ is between 5 and 10. At $\tan{\beta}< 5$ the LEP bound on $m_h$ is problematic. On the contrary, the higgsino-stop contribution to $b\rightarrow s+\gamma$ grows like $\tan{\beta}$ and, with mixing angles analogous to the CKM ones, becomes exceedingly large. As an example, this is illustrated in Fig. \ref{bsgamma}, which is well representative of the generic situation. While this is uncertain, also in view of possible cancellations with other contributions to the $b\rightarrow s+\gamma$ amplitude coming from charged Higgs or gluino exchanges, we find it better to stick to $\tan{\beta}< 10$. In fact the specific value taken by $\tan{\beta}$ between 5 and 10 constrains $m_{\tilde t_1}, m_{\tilde t_2}$ and $\theta_t$ through $m_h > 114$ GeV but is otherwise practically irrelevant in the branching ratios for the modes (\ref{sqchi}, \ref{sqV}, \ref{sqS}). Incidentally notice also from Fig.\ref{bsgamma}  that $b\rightarrow s+\gamma$ disfavours highly asymmetric stop masses because of the correspondingly growing stop-mixing parameter for fixed values of $\theta_t$. In the same $(m_{\tilde t_1}, m_{\tilde t_2})$-plane, on the contrary, there is no significant constraint from the $\rho$-parameter.

Largely dominant among the squark decays are the higgsino decays (\ref{sqchi}). 
This is a cumulative effect of phase space and of suppression factors of the decay amplitudes into the vector bosons and into scalars. Furthermore, among the same higgsino decays, one of them, $\tilde{b}\rightarrow b \chi$ also has a suppressed amplitude, relative to the other ones, by a factor $(m_b/m_t)\tan{\beta}$. This is because the coupling between $b_L$ and $b_R$ only occurs in the  down Yukawa  coupling. A good approximation for a general discussion, therefore, is provided by sticking to the following relevant decay modes:
\begin{equation}
\tilde{t}_{1,2} \rightarrow t \chi,~~
\tilde{t}_{1,2} \rightarrow b \chi^+,~~
\tilde{b} \rightarrow t \chi^-,
\end{equation}
and the corresponding conjugate ones. As already mentioned,  the decays $\tilde{t}_{1,2} \rightarrow t \chi$ are meant to be summed over the two neutral almost degenerate higgsinos. In turn,  the possible final states in the gluino chain decay are limited to:
\begin{equation}
\tilde{g}\rightarrow t \bar{t} \chi; ~~\tilde{g}\rightarrow t \bar{b} \chi^-;~~\tilde{g}\rightarrow  \bar{t} b \chi^+,
\end{equation}
with the last two conjugate decays having an identical branching ratio,
\begin{equation}
B_{tb}\equiv BR(\tilde{g}\rightarrow t \bar{b} \chi^-)= BR(\tilde{g}\rightarrow  \bar{t} b \chi^+) \approx \frac{1}{2}(1- BR(\tilde{g}\rightarrow t \bar{t} \chi)).
\label{defBtb}
\end{equation}
We find it a remarkable simplification that a single inclusive branching ratio may end up characterizing the main features of the phenomenology of 
this configuration of s-particle masses, as we are going to discuss. This remains true even if one wants to include in the gluino decay chain  the  subdominant modes $\tilde{q} \rightarrow \tilde{q} V$, $\tilde{q} \rightarrow \tilde{q} S$.
To this end one needs to generalize $B_{tb}$ to 
\begin{equation}
\bar{B}_{tb}\equiv BR(\tilde{g}\rightarrow t \bar{b} X^-)= BR(\tilde{g}\rightarrow  \bar{t} b X^+) 
\end{equation}
where $X^\pm =\chi^\pm$ or $\chi^\pm + h(Z)$ or $\chi_{1,2}+ W^\pm$. 
In any given point of the parameter space the difference between $B_{tb}$ and $\bar{B}_{tb}$ is small. Furthermore 
$\bar{B}_{tb}$  is equally useful to characterize the features of the events we study in the next Section.

\section{A single relevant semi-inclusive branching ratio}
\label{Other}

As it is clear from the previous considerations,  the most characteristic feature of the LHC signals from this configuration of s-particle masses is an overproduction of top quarks from gluino decays. Starting from gluino pair production, there will in fact be four types of semi-inclusive final states:
\begin{equation}
pp \rightarrow \tilde{g} \tilde{g}\rightarrow t t\bar{t} \bar{t} +\chi\chi
\label{4t}
\end{equation}
\begin{equation}
pp \rightarrow \tilde{g} \tilde{g}\rightarrow t t\bar{t} \bar{b}  (\bar{t} \bar{t} t b) +\chi\chi
\label{3t}
\end{equation}
\begin{equation}
pp \rightarrow \tilde{g} \tilde{g}\rightarrow t t\bar{b} \bar{b}  (\bar{t} \bar{t} b b) +\chi\chi
\label{2t}
\end{equation}
\begin{equation}
pp \rightarrow \tilde{g} \tilde{g}\rightarrow  t\bar{t} b\bar{b}   +\chi\chi
\end{equation}
where, as before, $\chi$  stands for a charged or any one of the two neutral higgsinos. Their rates will depend in good approximation, up to the overall  gluino-pair production rate, only on $B_{tb}$,
 \begin{equation}
R( t t\bar{t} \bar{t})=     B_{tt}^2, ~~B_{tt} \equiv BR(\tilde{g}\rightarrow t \bar{t} \chi) \approx 1-2 B_{tb}
\end{equation}
\begin{equation}
R( t t\bar{t} \bar{b}  (\bar{t} \bar{t} t b) )=     4 B_{tt} B_{tb}
\end{equation}
\begin{equation}
R( t t\bar{b} \bar{b}  (\bar{t} \bar{t} b b) )  =    2 ( B_{tb})^2
\end{equation}
\begin{equation}
R( t\bar{t} b\bar{b})=    2 ( B_{tb})^2.
\end{equation}
$B_{tb}$ is therefore the crucial quantity to study.

Whenever $m_t$ is negligible in  phase space factors, it is $B_{tb}\approx 25\%$. This follows from the form of the top Yukawa coupling, which, as already mentioned, dominates over the bottom one and dictates the couplings of the left and right squarks to the higgsinos
\begin{equation}
\Delta \mathcal{L} = -\lambda_t (  \tilde t_R^*\overline\chi P_L t + \tilde t_L^* \overline\chi P_R t -  \tilde t_R^*\overline{\chi^-} P_L b - \tilde b_L^* \overline{\chi^+}P_R t) +h.c.
\label{yukawa-stops}
\end{equation}
where both the charged $\chi^\pm$ and the neutral $\chi$, in the limit of degenerate $\chi_1, \chi_2$, can be considered as Dirac spinors of the same mass $\mu$.

Neglect first $\tilde{t}_L-\tilde{t}_R$ mixing. Then from (\ref{yukawa-stops}) and $m_{\tilde{t}_L} \approx m_{\tilde{b}}$ (up to $m_t$ corrections)
\begin{equation}
BR(\tilde{g}\rightarrow \tilde{t}_L \bar{t}\rightarrow t \bar{t} \chi)\approx  BR(\tilde{g}\rightarrow \tilde{b}_L \bar{b}\rightarrow t \bar{b} \chi) 
\label{gluino_tL}
\end{equation}
and
\begin{equation}
BR(\tilde{g}\rightarrow \tilde{t}_R \bar{t}\rightarrow t \bar{t} \chi)\approx  BR(\tilde{g}\rightarrow \tilde{t}_R \bar{t}\rightarrow b \bar{t} \chi) 
\label{gluino_tR}
\end{equation}
with the same branching ratios for the charge conjugate modes. Hence indeed $B_{tb} \approx 25\%$, either from (\ref{gluino_tL}) or from (\ref{gluino_tR}), which does not change by $\tilde{t}_L-\tilde{t}_R$ mixing since $B_{tb}$ is inclusive. On the other hand, whenever $m_t$ becomes important in phase space factors, $B_{tb}$ increases since the $b$-reach modes are favoured over the $t$-reach ones.

\begin{figure}[tb]\label{Btb}
\centering
\includegraphics[width=8cm]{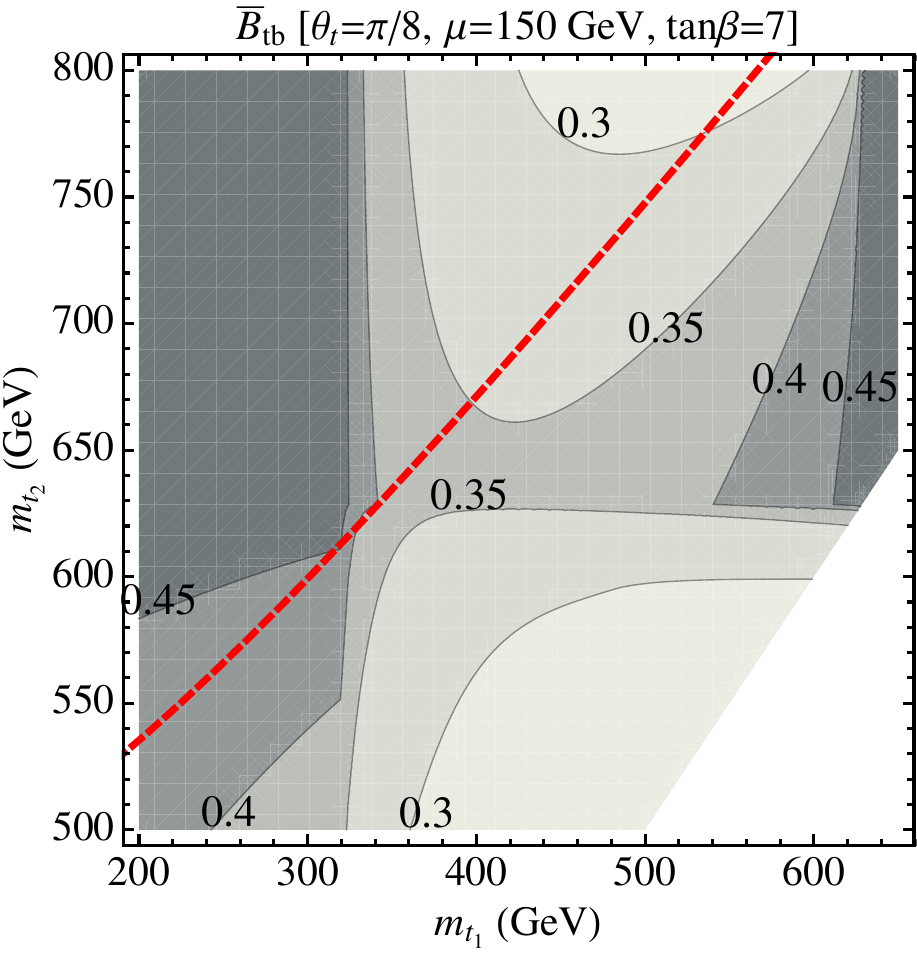} \includegraphics[width=8cm]{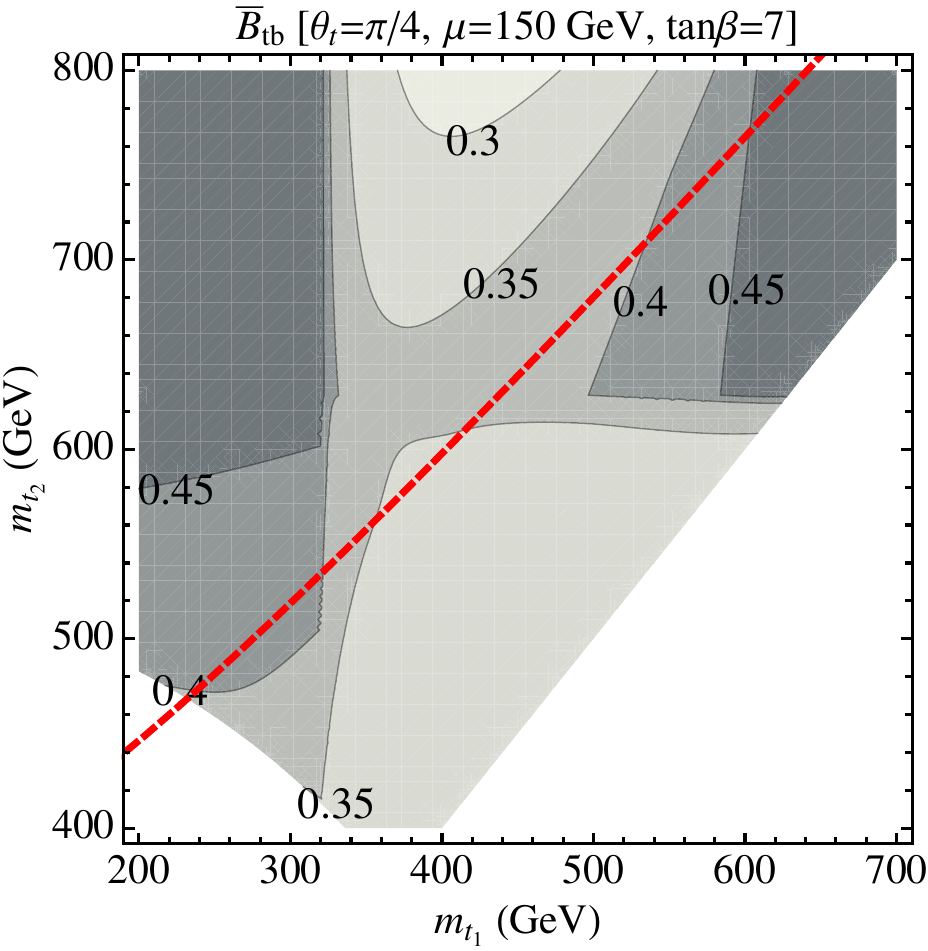} \caption{{\small Contour lines for $\bar B_{tb}$ as defined in the text, for $m_{\tilde g}= 800\,\textrm{GeV}$, $\tan\beta=7$, $\mu=150\,\textrm{GeV}$ and $\theta_t=\pi/8,\,\pi/4$.  The Higgs boson mass satisfies $m_h>114\,\textrm{GeV}$ (see eq. (\ref{higgsmass}) in the Appendix)  above the red dashed line.}}
\end{figure}

This is confirmed and illustrated in Fig.\ref{Btb} for some sample values of $m_{\tilde{g}}, \mu, \theta_t$, where the bound on the Higgs mass from LEP is also shown. These Figures all exhibit the same pattern, which is in particular almost symmetric in $\theta_t$ above or below $\pi/4$.  From $25\%$ $B_{tb}$ increases and eventually reaches its limiting value $B_{tb}\approx 50\%$ for asymmetric physical stop masses, where $\tilde{t}_1\rightarrow t \chi$ gets inhibited by phase space, unlike $\tilde{t}_1\rightarrow b \chi$, or when both $\tilde{t}_1$ and $\tilde{t}_2$ are heavy enough that $\tilde{g}\rightarrow \tilde{t}_{1,2}t$ gets kinematically constrained or even forbidden, unlike $\tilde{g}\rightarrow \tilde{b} b$.

\section{Same sign di-leptons and tri-leptons from top decays}

\begin{figure}[tb]
\label{rates}
\centering
\includegraphics[width=9cm]{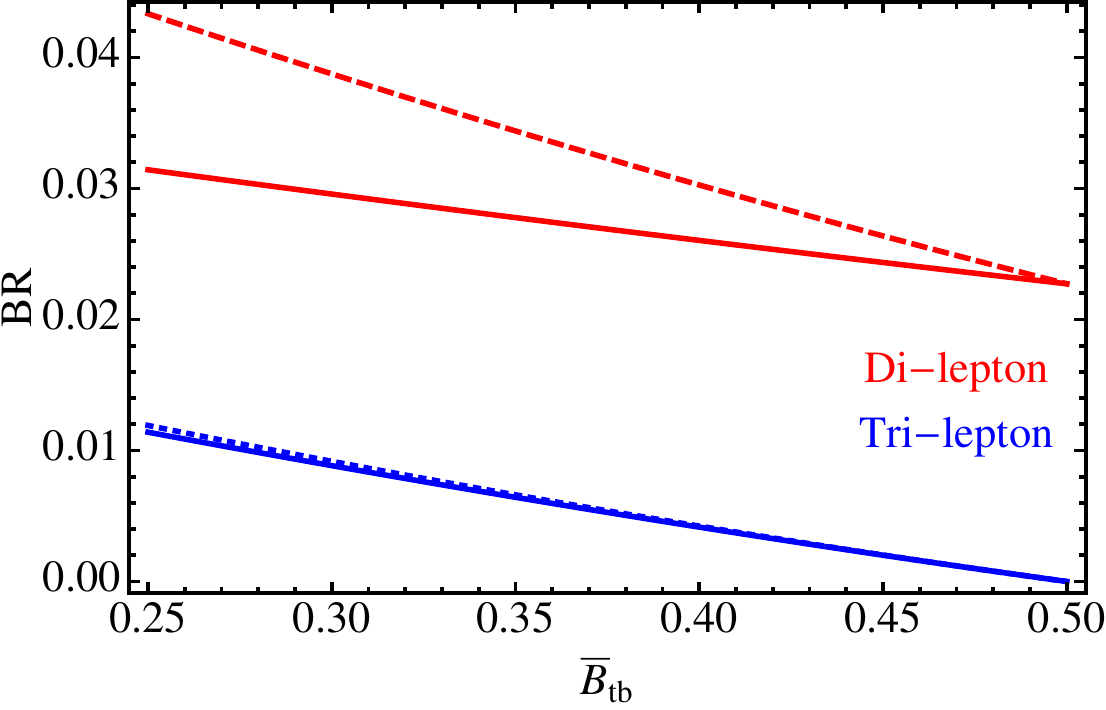}\caption{{\small Rates for the same-sign di-lepton signal (red line) and tri-lepton signal (blue line) as a function of $\bar B_{tb}$. The red (blue) dashed line indicates the rate to have \emph{at least} two same-sign leptons (three leptons).}}
\end{figure}

High $p_T$ same-sign di-lepton and tri-lepton events both from supersymmetry\footnote{Early references include \cite{Gamberini:1986eg} \cite{Baer:1986au} \cite{Barbieri:1991vk} \cite{Barnett:1993ea}\cite{Guchait:1994zk}, whereas for more recent ones see \cite{Hisano:2002xq} \cite{Kraml:2005kb}\cite{Martin:2008aw}. Multi-lepton events specifically generated from semi-leptonic top decays have also been considered in ref.s \cite{Toharia:2005gm}\cite{Mercadante:2005vx}\cite{Baer:2007ya}\cite{Acharya:2009gb}, although not originating from the s-particle configuration considered here.} as from other Beyond-the-Standard-Model physics\cite{Lillie:2007hd}\cite{Contino:2008hi}\cite{Pomarol:2008bh}, have been extensively discussed in the literature. They look particularly relevant in the present case,  coming from the semi-leptonic decays of the abundant top quarks. Adding up the same-sign di-leptons, either electrons or muons, from (\ref{4t},\ref{3t},\ref{2t}), one finds a total rate, for any gluino pair produced,
\begin{equation}
R(l^\pm l^\pm + jets + E_{Tmiss}) = 2 B_l^2 (B_{tb} + (1-2 B_{tb}) B_h)^2
\end{equation}
where $B_l = 21\%$ and $B_h = 68 \% $ are the branching ratios of the $W$ into $e, \mu$ or into hadrons respectively. Each of these events contains four $b$-jets and up to four extra partonic jets from $W$ decays, together with the missing energy associated with the two $\chi$'s and two neutrinos from the top quarks. Similarly the semi-leptonic decay of three top quarks from either (\ref{4t}) or (\ref{3t}) give a rate of tri-lepton events
\begin{equation}
R(l^\pm l^+ l^- + jets + E_{Tmiss}) = 4 B_l^3 (1-2 B_{tb}) (B_{tb} + (1-2 B_{tb}) B_h)
\end{equation}
where this time, other than the four $b$-jets, there are up to two additional jets.
These relative rates are shown in Fig.\ref{rates} versus $B_{tb}$.

\begin{table}[tb]\label{tabella}
\centering
\begin{tabular}{c|c|c|c}
BKG & Total (LO) (fb) & Di-lepton (SS)& Tri-lepton\\
\hline
$t\overline t$ & 442 (pb) & 400 & - \\
$t\overline t W^\pm$ & 465 & 15 & 5\\
$t\overline t W^-W^+$ & 9 & 0.4 & 0.2\\
$W^+W^-$& 73 (pb)& 66& - \\
$W^\pm W^\pm(jj)$& 480 & 20& - \\
$W^+W^-W^\pm$& 130&  4 & 1\\
\hline
\end{tabular}
\caption{{\small Leading-order cross sections for the main backgrounds to the signal. The same-sign di-lepton cross section is estimated including the possibility to misidentify the lepton charge with a 1\% probability in the $t\bar{t}$ and in the $W^+W^-$ cases.
}}
\end{table}

These signals should allow the discovery of the configuration of s-particles that we are discussing already in the relatively early stages of LHC. Taking, e.g., $m_{\tilde{g}}= 800$ GeV and $B_{tb} = 0.3$, the cross section at $\sqrt{s}= 14$ TeV for the same sign di-leptons (tri-leptons) from top decays is $23 fb (7.5 fb)$. This should be compared with the background raw rates estimated in Table 1, without the imposition of any cut\footnote{In our analysis we use PYTHIA\cite{pythia} to generate signal events and MADGRAPH\cite{madgraph} for the backgrounds.}. 

\begin{figure}[tb]
\centering
\includegraphics[width=9cm]{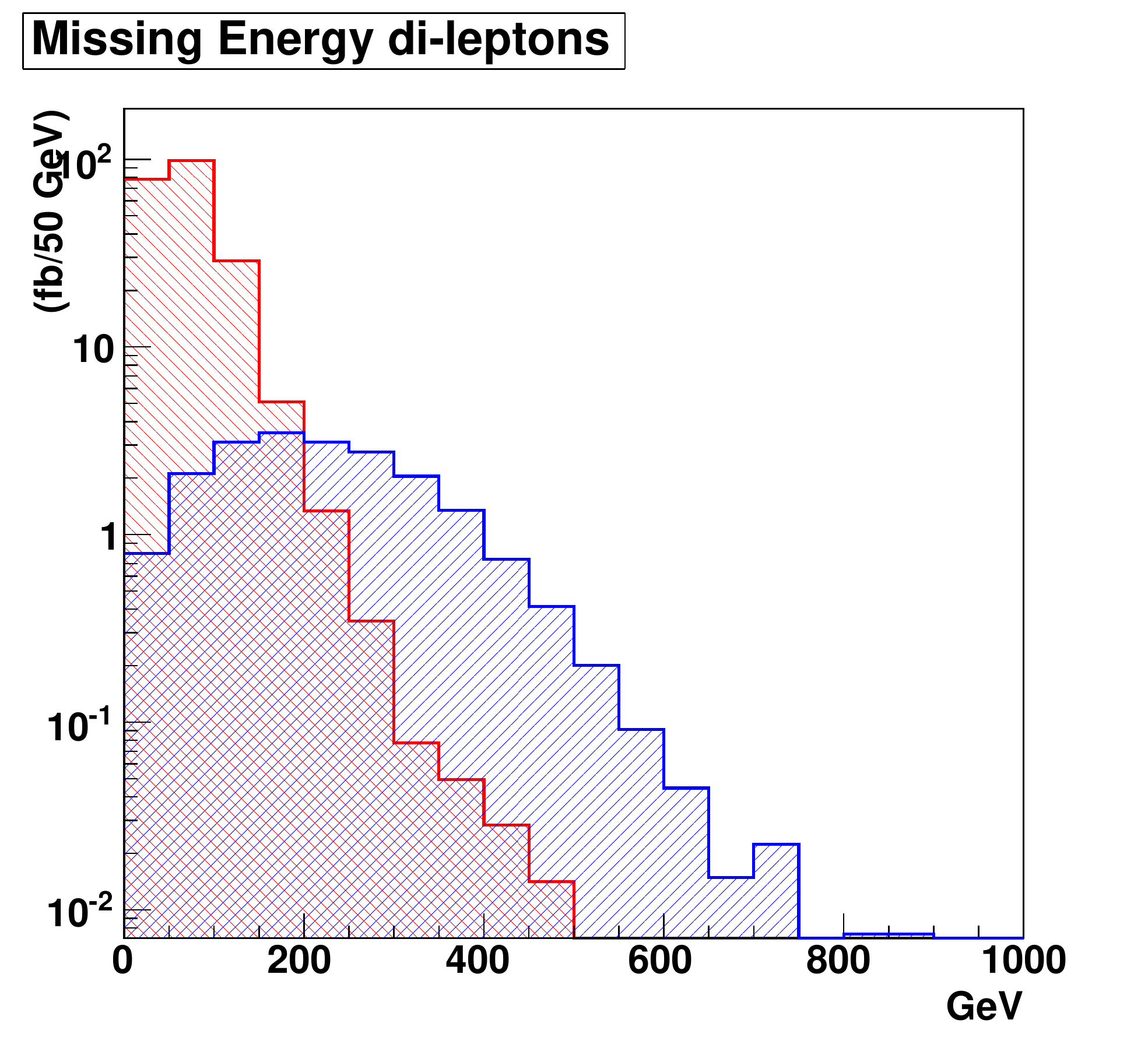}\includegraphics[width=9cm]{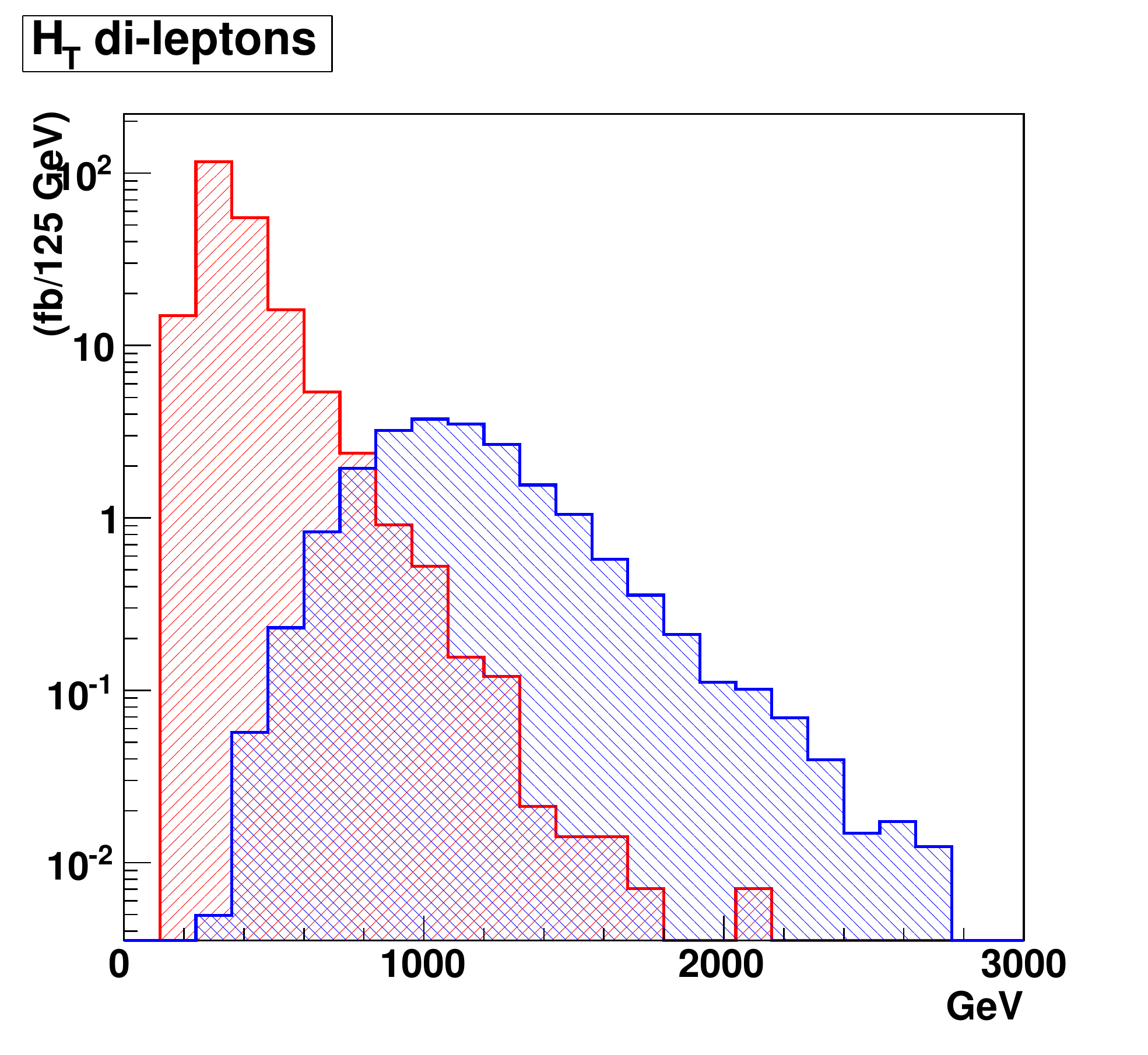} \caption{{\small Distributions of the missing transverse energy (left) and of $H_t$ (right) for the signal events where at least two same-sign leptons with $p_T>30\textrm{GeV}$ are found. Also shown are the same distributions from the dominant $t\bar{t}$ background (see text).}}
\label{dist}
\end{figure}

The two dominant backgrounds exceeding the signal arise from $t\bar{t}$ or $W^+ W^-$ with one of the leptons  mis-identified to have the wrong charge with a $1\%$ probability. Fig. \ref{dist} shows some distributions for the same sign di-lepton signal that should allow to discriminate it  relatively easily from these  backgrounds. Particularly effective would be a cut on the scalar sum, $H_t$, of all the transverse momenta and of the transverse missing energy, that would only marginally affect the signal. In Fig. \ref{dist} we have taken $m_{\tilde{g}}= 800$ GeV, $m_{\tilde{t}_1} = 400$ GeV,  $m_{\tilde{t}_2} = 680$ GeV, $\theta_t = \pi/8$ and $\mu= 150$ GeV, corresponding to $B_{tb} \approx 0.35$. To see a possible effect of the leptons in the cascade decays among the higgsinos we have also introduced a splitting among them by mixing with the electroweak gauginos at $M_1 = 600$ GeV and $M_2 = 500$ GeV.
Before any cuts are applied, the cross section at $\sqrt{s}= 14$ TeV for the same sign di-leptons (tri-leptons) from top decays is $21\, fb\, (5\, fb)$. 
With an integrated luminosity of $1 fb^{-1}$, cutting at $H_t = 1$ TeV leaves about 14 events with at least two same-sign leptons with $p_T>30\textrm{GeV}$ 
against less than one event of our simulated total background\footnote{In roughly 10 of these 14 events we checked that the same-sign di-leptons (tri-leptons) truly originate from the decay of top quarks.}. No attempt is made to optimize the cuts, since a more realistic description of the background may be required.

\section{Summary and possible extensions}

A problem one encounters in studying the phenomenology of the MSSM is the huge number of parameters that plagues the most general version of it. Even specific theoretically motivated versions of the MSSM, based on particular mechanisms of supersymmetry breaking and transmission, involve quite a number of parameters, making difficult to give a compact description of the relevant phenomenology. This is at least one good reason for looking at a particular configuration of s-particle masses that appears  theoretically motivated and has the advantage of being describable  in terms of relatively few physical parameters. The typical spectrum of one such configuration of s-particles is  shown in Fig. \ref{spettro}, with the s-particle masses allowed to vary in the range given in Sect. 1. No doubt this is a particular case of the generic MSSM. Yet it is useful  to concentrate one's attention on it - we believe - as a particularly interesting case, for the reasons also given in Sect. 1.

The most characteristic signal to be expected at the LHC is due to the relative abundance of multi-top events. There may be several ways to attack the detection of this signal. Most significantly, however, the top decays will produce a number of hard $p_T$ same-sign di-leptons or tri-lepton events with rates, for any given gluino mass,  that can be easily read from Fig.s \ref{Btb} and \ref{rates} in any point of the parameter space. What counts most is the inclusive branching ratio $B_{tb}$ defined in eq. (\ref{defBtb}). Although the study of the precise discovery potential of the LHC has not been our main motivation, we believe that it should be possible to explore the entire range of masses described in Sect. 1 already in a relatively early stage of the LHC, say at $\sqrt{s}=14$  TeV and an integrated luminosity of 1 $fb^{-1}$.
We would not be too surprised if this turned out to be the first signal of manifest supersymmetry.

How generic are these considerations from the point of view of the LHC signals?  While they are certainly not completely generic, they can apply with little or even no modifications at all also to some other configurations of s-particle masses, more or less theoretically motivated but certainly possible.   The important thing is that no squarks other than $\tilde{t}_1, \tilde{t}_2$ and $\tilde{b}$ are introduced below 1 TeV. Sleptons, on the contrary, can be anywhere above $\mu$. These modifications alone would introduce essentially no change of the picture described above. Some change would on the contrary occur by bringing down the electroweak gaugino masses relative to Fig. \ref{spettro}. This would make the full chargino-neutralino sector enter into the gluino decay chain, with the net result, however, of increasing the number of high $p_T$ multi-lepton events. Other than those ones coming from the semi-leptonic top decays, one would also have the leptons from the cascade decays
inside the chargino-neutralino sector. This is partly true even in the case of the spectrum of Fig. \ref{spettro}, depending on the cut of the  $p_T$ of the leptons. A motivation for looking at some modifications of the s-particle spectrum of Fig. \ref{spettro} along the lines mentioned above might come from Dark Matter considerations. If thermally produced in the early universe, the cosmological abundance of the lightest higssino in Fig. \ref{spettro} would in fact be insufficient to make the observed amount of Dark Matter.

Finally one can ask what happens on the low side of the gluino mass range that we have considered, or even below it, for the same configuration of the other s-particles. The answer is clear. As long as $\tilde{g}\rightarrow  t \bar{b} \chi$ is kinematically allowed to go, there would again be no significant modification of the picture described above with $B_{tb}$ close to 0.5. At extreme values of the parameters, - say $\mu \approx 100$ GeV and $m_{\tilde{g}} = 300\div 350$ GeV -, this might even be of interest for the current searches at the Tevatron\cite{Tevatrone}.

\section*{{Acknowledgments}}

{This work is supported by the EU under RTN
contract MRTN-CT-2004-503369, by the
MIUR under contract 2006022501 and by the Swiss National Science Foundation under contract No. 200021-116372. We thank Gino Isidori
for many 
useful discussions and comments; D.P. thanks Roberto Franceschini and Jan Mrazek for valuable hints and discussions.}

\appendix

\section{Appendix}
\subsection{Gluino and s-quarks partial decay widths}
For ease of the reader we collect here the expressions for the relevant decay widths used in the text. In each partial amplitude the same phase-space factor appears, so we define 
\begin{equation}
\Phi(x_1,x_2)=\sqrt{\left(1-(x_1-x_2)^2\right)\left(1-(x_1+x_2)^2\right)}.
\end{equation}
The two-body gluino decay amplitudes are given by
\begin{eqnarray}
&&\Gamma(\tilde g\rightarrow\tilde t_{1,2} t)=\alpha_S \frac{m_{\tilde g}}{8}\left(1-\frac{m_{\tilde t^2_{1,2}}}{m^2_{\tilde g}}+\frac{m^2_t}{m^2_{\tilde g}}\mp2\frac{m_t}{m_{\tilde g}}\sin2\theta_t\right)\Phi\left(\frac{m_{\tilde t_{1,2}}}{m_{\tilde g}},\,\frac{m_t}{m_{\tilde g}}\right)\\
&&\Gamma(\tilde g\rightarrow\tilde b_{1} b)=\alpha_S \frac{m_{\tilde g}}{8}\left(1-\frac{m_{\tilde b^2_{1}}}{m^2_{\tilde g}}+\frac{m^2_b}{m^2_{\tilde g}}\right)\Phi\left(\frac{m_{\tilde b_{1}}}{m_{\tilde g}},\,\frac{m_b}{m_{\tilde g}}\right).
\end{eqnarray}
Finally we list the partial widths for the \emph{stops} and the \emph{sbottom} s-quarks
\begin{eqnarray}
&&\Gamma(\tilde t_{1,2}\rightarrow t\chi)= \frac{m_{\tilde t_{1,2}}}{32\pi}\left(\frac{m_t}{v \sin\beta}\right)^2\left(1-\frac{m_t^2}{m^2_{\tilde t_{1,2}}}-\frac{m_\chi^2}{m^2_{\tilde t_{1,2}}}\pm2\frac{m_t m_\chi}{m^2_{\tilde t_{1,2}}}\sin2\theta_t\right)\cdot\\
&&\hspace{10.5cm}\cdot\,\Phi\left(\frac{m_t}{m_{\tilde t_{1,2}}},\,\frac{m_\chi}{m_{\tilde t_{1,2}}}\right)\\
&&\Gamma(\tilde t_{1}\rightarrow b\chi^+)= \frac{m_{\tilde t_1}}{16\pi}\left(\frac{m_t}{v \sin\beta}\right)^2\left(1-\frac{m_b^2}{m^2_{\tilde t_{1}}}-\frac{m_\chi^2}{m^2_{\tilde t_{1}}}\right)\cos^2\theta_t\Phi\left(\frac{m_b}{m_{\tilde t_{1}}},\,\frac{m_\chi}{m_{\tilde t_{1}}}\right)\\
&&\Gamma(\tilde t_{2}\rightarrow b\chi^+)= \frac{m_{\tilde t_{2}}}{16\pi}\left(\frac{m_t}{v \sin\beta}\right)^2\left(1-\frac{m_b^2}{m^2_{\tilde t_{2}}}-\frac{m_\chi^2}{m^2_{\tilde t_{2}}}\right)\sin^2\theta_t\Phi\left(\frac{m_b}{m_{\tilde t_{2}}},\,\frac{m_\chi}{m_{\tilde t_{2}}}\right)\\
&&\Gamma(\tilde b_{1}\rightarrow t\chi^-)= \frac{m_{\tilde b_1}}{16\pi}\left(\frac{m_t}{v \sin\beta}\right)^2\left(1-\frac{m_t^2}{m^2_{\tilde b_{1}}}-\frac{m_\chi^2}{m^2_{\tilde b_{1}}}\right)\Phi\left(\frac{m_b}{m_{\tilde b_1}},\,\frac{m_\chi}{m_{\tilde b_1}}\right)\\
&&\Gamma(\tilde t_2\rightarrow \tilde t_1 Z)=\frac{\alpha}{s_W c_W}\frac{m_{\tilde t_2}}{32}\left(-1-\frac{m^2_{\tilde t_1}}{m^2_{\tilde t_2}}+\frac{m^2_{\tilde t_2}}{2M_Z^2}\left(1-\frac{m^2_{\tilde t_1}}{m^2_{\tilde t_2}}\right)^2+\frac{M_Z^2}{2m^2_{\tilde t_2}}\right)\cdot\\
&&\hspace{9.5cm}\cdot\sin^22\theta_t\,\Phi\left(\frac {M_Z}{m_{\tilde t_2}},\,\frac{m_{\tilde t_1}}{m_{\tilde t_2}}\right)
\end{eqnarray}
\begin{eqnarray}
&&\Gamma(\tilde t_2\rightarrow \tilde b_1 W)=\sqrt 2 G_F M_W^2\frac{m_{\tilde t_2}}{8\pi}\left(-1-\frac{m^2_{\tilde b_1}}{m^2_{\tilde t_2}}+\frac{m^2_{\tilde t_2}}{2M_W^2}\left(1-\frac{m^2_{\tilde b_1}}{m^2_{\tilde t_2}}\right)^2+\frac{M_W^2}{2m^2_{\tilde t_2}}\right)\cdot\\
&&\hspace{9.5cm}\cdot\cos^2\theta_t\,\Phi\left(\frac{M_W}{m_{\tilde t_2}},\,\frac{m_{\tilde b_1}}{m_{\tilde t_2}}\right)\\
&&\Gamma(\tilde t_2\rightarrow \tilde t_1 h)=\sqrt 2 G_F M_W^2\frac{m^3_{\tilde t_2}}{64\pi}\left(1-\frac{m^2_{\tilde t_1}}{m^2_{\tilde t_2}}\right)\cos^22\theta_t\sin^22\theta_t\,\Phi\left(\frac{m_{\tilde b}}{m_{\tilde t_2}},\,\frac{m_{h}}{m_{\tilde t_2}}\right).
\end{eqnarray}

\subsection{The radiatively corrected Higgs boson mass}
We quote the formula used for the calculation of the Higgs boson mass. This accounts for the one-loop and the leading-log two loops contribution from the top and the stops \cite{Carena:1995bx}.
\begin{equation}
m_h^2=M_Z^2 \cos^22\beta\left(1-\frac{3}{8\pi^2}\frac{M_t^2}{v^2}\log\frac{M_S^2}{M_Z^2}\right)+
\label{higgsmass}
\end{equation}
\begin{equation*}
+\frac{3}{4\pi^2}\frac{M_t^4}{v^4}\left(\frac{X_t}{2}+\log\frac{M_S^2}{M_Z^2}+\frac{1}{16\pi^2}\left(\frac{3}{2}\frac{M_t^2}{v^2}-32\pi^2\alpha_S(M_t)\right)\left(X_t\log\frac{M_S^2}{M_Z^2}+\log^2\frac{M_S^2}{M_Z^2}\right)\right)
\end{equation*}
with
\begin{equation}
X_t=2\frac{\widetilde A^2_t}{M_S^2}\left(1-\frac{\widetilde A^2_t}{12 M_S^2}\right),\qquad \widetilde A_t=A_t-\frac{\mu}{\tan\beta}=\frac{m_{\widetilde t_2}^2-m_{\widetilde t_1}^2}{2 M_t}\sin 2\theta_t
\end{equation}
and
\begin{equation}
M_S^2=\frac{m_{\widetilde t_1}^2+m_{\widetilde t_2}^2}{2},
\end{equation}
where $M_t$ is the running top quark mass evaluated at the pole.

\end{document}